\documentclass[twocolumn,floatfix,prl,aps,showpacs]{revtex4-1}
\usepackage{graphicx,amsmath,amssymb}

\begin{document}

\newcommand{\CORR}[1]{\textbf{#1}}

\title{Fractional quantum Hall effect arising from repulsive three body interaction}

\author{Arkadiusz W\'ojs$^{1,2}$, Csaba T\H oke$^{3}$, and Jainendra K.~Jain$^{4}$}

\affiliation{$^{1}$TCM Group, Cavendish Laboratory, University of Cambridge, Cambridge CB3 0HE, United Kingdom}

\affiliation{$^{2}$Institute of Physics, Wroclaw University of Technology, 50-370 Wroclaw, Poland}

\affiliation{$^3$Institute of Physics, University of P\'ecs, 7624 P\'ecs, Hungary}

\affiliation{$^{4}$Department of Physics, 104 Davey Lab, Pennsylvania State University, University Park PA, 16802}

\date{\today}

\begin{abstract}
We consider a collection of fermions in a strong magnetic field coupled by a purely three body repulsive interaction, and predict the formation of composite fermions, leading to a remarkably rich phase diagram containing a host of fractional quantum Hall states, a composite fermion Fermi sea, and a pairing transition.  This is entirely unexpected, because the appearance of composite fermions and fractional quantum Hall effect is ordinarily thought to be a result of strong two-body repulsion. Recent theoretical and experimental breakthroughs in ultra-cold atoms and molecules have facilitated the realization of such a system, where this physics can be tested.
\end{abstract}


\maketitle

The fractional quantum Hall effect \cite{Tsui82} (FQHE) occurs as a result of the repulsive Coulomb interaction when two-dimensional electrons are subjected to a strong magnetic field. The nature of the emergent state is dependent on the form of the interaction and it is natural to wonder what physics would be produced by other types of interactions. We consider in this Letter the system of fermions in the lowest Landau level (LL) interacting via a purely 3-body interaction.  The original motivation for considering such an interaction \cite{Greiter91} was the observation that the Pfaffian wave function \cite{Moore91}, which describes a paired state of composite fermions (CFs), is obtained as the exact solution at LL filling factor $\nu=1/2$ when only the first nontrivial pseudopotential of the 3-body interaction is retained. This model also produces exact solutions for quasiholes, which are believed to exhibit nonabelian braid statistics \cite{Moore91,Read00}. 

Model interactions that are truncated beyond a finite number of interaction pseudopotentials have zero energy solutions that can be explicitly enumerated. (The pseudopotential $V^{(n)}_m$ is defined \cite{Haldane83} as the energy of the collection of $n$ particles in a state with relative angular momentum $m$.) However, they have much physics not captured by the zero energy solutions. Take, for example, the 2-body interaction wherein all pseudopotentials except $V^{(2)}_1$ are set to zero. The zero energy solutions occur for $\nu\leq 1/3$, but this interaction also produces extensive phenomenology for $\nu\geq 1/3$. The more complete physics of this model interaction (and also of the Coulomb interaction) is captured by the CF theory \cite{Jain89}, which gives a qualitative understanding of incompressibility at $\nu=n/(2n\pm 1)$, and also very accurate solutions for the incompressible ground states, their neutral excitations, quasiholes, and quasiparticles. 

Our aim in this paper is to explore the general physics of the 3-body repulsive  interaction.  We find that for a large range of parameters the physics of the 3-body interaction is well described by composite fermions, resulting in almost as extensive FQHE as that seen for electrons in GaAs (albeit with with crucial  differences).  This is surprising because composite fermions and the FQHE are widely believed to be caused by a strong repulsion in the two body channel.  

Three body interaction arises in FQHE system due to LL mixing, which breaks particle hole symmetry; it plays a role at $\nu=5/2$ in lifting the degeneracy between the Pfaffian and its hole conjugate \cite{Levin07}. In the typical FQHE systems, however, it is only a small perturbation to the 2-body Coulomb interaction. A model that has {\em only} 3-body interaction might therefore appear physically irrelevant at first sight.  However, significant strides have been made toward the realization of such interaction in ultra-cold atomic or molecular systems.  On the one hand, synthetic magnetic fields have been demonstrated by rapid rotation \cite{Schweikhard04} and by inducing a Berry phase through a nonuniform coupling between the internal states \cite{Lin09}. Additionally, ingenious proposals have been advanced for eliminating the pairwise interaction by tuning external parameters \cite{Cooper04}, leaving the three body interaction as the dominant term. It is therefore possible that the phase diagram predicted in this work can be experimentally accessed in ultra-cold atoms.  

Of relevance to what follows are composite fermions \cite{Jain89}, topological bound states of electrons and an even number ($2p$) of quantized vortices. As a result of the bound vortices, composite fermions experience a reduced effective magnetic field ${\cal B}^*={\cal B}-2p\rho hc/e$, where ${\cal B}$ is the external magnetic field and $\rho$ is the particle density; in particular, composite fermions form $\Lambda$ levels analogous to the LLs at ${\cal B}^*$.  Much physics can be understood by neglecting the interaction between composite fermions.  Composite fermions' integral quantum Hall effect (IQHE) manifests as the FQHE at $\nu=n^*/(2pn^*\pm 1)$ ($n^*$ is an integer); their excitons are the lowest energy neutral excitations \cite{Kang01,Dev92}; and their Fermi sea describes the compressible state at the half filled LL \cite{Halperin93,Willett93}. In certain cases, the residual interaction between the composite fermions plays an important role. Certain fractions such as 4/11 represent a FQHE of composite fermions \cite{Pan03,Chang04}.  At $\nu=5/2$, a residual attraction between composite fermions \cite{Scarola00} causes their $p_x\pm ip_y$ pairing, described by the above mentioned Pfaffian wave function \cite{Moore91,Greiter91,Read00}. 

We consider fully spin polarized fermions in the lowest LL, subject to a model three body interaction in which all but the first two relevant pseudopotentials, $V^{(3)}_{3}\equiv A$ and $V^{(3)}_{5}\equiv B$, are set to zero.  (For three body interaction, $m=1$ and $2$ are excluded by the Pauli principle and $m=4$ by symmetry.)  The most reliable method available for determining the states produced by a general interaction is that of exact diagonalization, for which we employ a geometry \cite{Haldane83} that has $N$ fermions moving on the surface of a sphere, subjected to a total radial magnetic flux of $2Q$ (in units of $hc/e$). The projected Hamiltonian assumes the form \cite{Simon07a}
\begin{equation}
\label{hami}
\hat H_{\rm 3-body}=A\sum_{i<j<k}P^{(3)}_{ijk}(3Q-3) + B\sum_{i<j<k}P^{(3)}_{ijk}(3Q-5),
\end{equation}
where $P^{(3)}_{ijk}(L)$ projects the state of the three particles $(i,j,k)$ into the subspace of total orbital angular momentum $L$. This Hamiltonian admits an exact solution for the ground state at $\nu=1/2$ (the Pfaffian) when $B=0$, and for the ground state at $\nu=2/5$ when both $A$ and $B$ are nonzero \cite{Simon07b}. We consider below the region $\nu>2/5$; there is no FQHE for this model for $\nu<2/5$. Increasing $B/A$ amounts to extending the range of the interaction. 

\begin{figure}
\begin{center}
\includegraphics[width=\columnwidth]{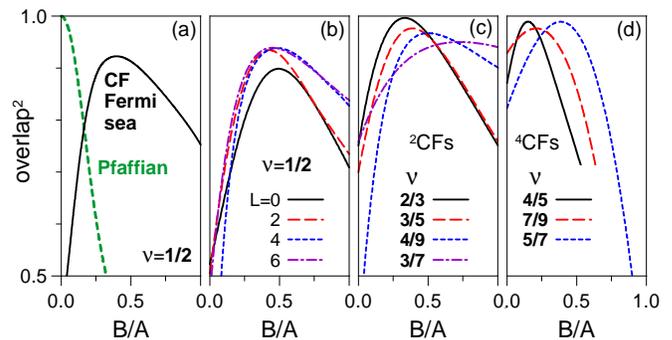}
\end{center}
\caption{\label{overlap} (Color online) Evolution of various states as a function of $B/A$, i.e., the range of the 3-body interaction. Panel (a) shows the squared overlap of the lowest energy $L=0$ eigenfunction of $H_{\rm 3-body}$ (which is also the ground state for $B/A<0.3$) with the Pfaffian wave function as well as with the $L=0$ state of weakly interacting  composite fermions at $\nu=1/2$ ($N\!=\!18$, $2Q\!=\!33$). The spherical geometry is used for these calculations, and $L$ is the total orbital angular momentum.  Panel (b) shows the squared overlaps of all CF states in the lowest energy band with the corresponding lowest energy eigenstates of $H_{\rm 3-body}$ for 14 fermions at $2Q=25$.  Panels (c) and (d) display comparison of the exact ground states of $H_{\rm 3-body}$ at $\nu=2/3$, 3/5, 4/9, 3/7, 4/5, 7/9 and 5/7 with the corresponding wave functions of IQHE states of composite fermions (for $N=$ 22, 18, 16, 15, 32, 30 and 24, respectively). 
} 
\end{figure}

\begin{figure*}
\begin{center}
\includegraphics[width=1.8383\columnwidth]{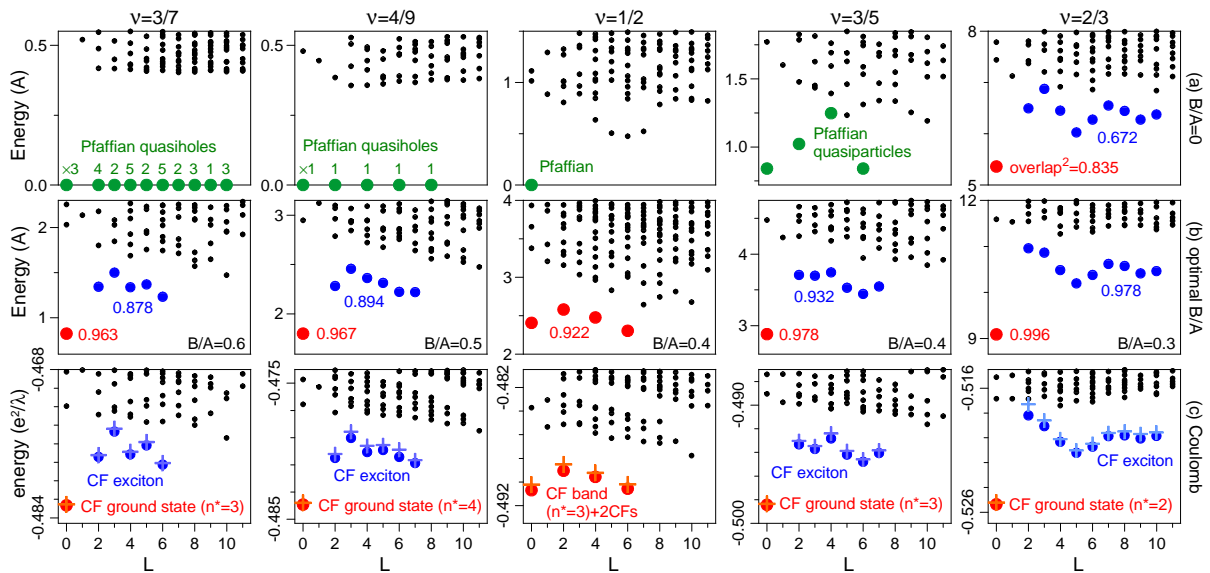}
\end{center}
\caption{\label{spectra} (Color online) Seeing composite fermions in the energy spectra for three body interaction at several filling factors shown at the top.  The top row (a) shows the spectra of $H_{\rm 3-body}$ for $B/A=0$, and the middle row (b) at the ``optimal value" of $B/A$ (see text for definition).  The spectra in the bottom row (c) are for the Coulomb interaction, as obtained by exact diagonalization (dots) and  from the CF theory (pluses) without any adjustable parameters.  The results at $\nu=$ 3/7, 4/9, 1/2, 3/5 and 2/3 are for 12, 16, 14, 15 and 18 particles, respectively. The Pfaffian ground state, quasiholes and quasiparticles are shown in green, with the integers near each Pfaffian quasihole dot indicating its degeneracy; the CF ground states are marked in red, and the CF excitons in blue. The squared overlaps with the corresponding CF states are indicated in the top two rows whenever the qualitative similarity with the CF spectrum warrants a comparison; to avoid clutter, only the average squared overlap is given for the exciton branch and also for the low-energy band at $\nu=1/2$. The energy $E$ is quoted in units of $A$ in the top two rows and $e^2/\lambda$ in the bottom row, where $\lambda=\sqrt{\hbar c/e{\cal B}}$ is the magnetic length; for the Coulomb interaction the quoted energy is per particle, and includes the electron-electron, electron-background, and background-background contributions. The x-axis label $L$ is the total orbital angular momentum, a good quantum number for the spherical geometry used in our calculations. 
}
\end{figure*}

We begin by testing the evolution of the Pfaffian solution as a function of the range of the interaction.   The overlap of the exact ground state of $\hat H_{\rm 3-body}$ with the Pfaffian wave function is seen in Fig.~\ref{overlap}(a) to decay rapidly with increasing $B/A$, indicating a transition into some other state. To gain insight into the nature of the new state we monitor the low energy spectrum, which reveals a fundamental restructuring as a function of $B/A$: As seen in the middle column of Fig.~\ref{spectra}, a distinct low energy band forms by the time the interaction reaches $B/A\sim 0.2$, and persists all the way up to $B/A=1$ and even beyond.  The crucial clue comes from the observation that this band has a perfect resemblance to the low energy band of almost free composite fermions at the effective flux $2Q^*=2Q-2(N-1)=-1$ (Fig.~\ref{spectra}).  The presence of such a band is a clear signature of the formation of weakly interacting composite fermions.  Further corroboration comes from a direct comparison of the exact eigenstates with the wave functions of weakly interacting  composite fermions. The latter can be obtained in the standard manner \cite{Jain89,Jain97} by ``composite-fermionizing" the known wave functions of noninteracting fermions at $2Q^*$. For the present work, we define the {\em Coulomb} eigenfunctions as the wave functions of weakly interacting composite fermions; the two are known to be practically identical \cite{Dev92,Jain97}, as also evident by an almost exact agreement between their Coulomb energies (Fig.~\ref{spectra}, lowest panels). Figs.~\ref{overlap}(a) and (b) demonstrate that as the overlap with the Pfaffian wave function drops, the overlap with the weakly interacting CF wave functions rapidly grows, approaching a high maximum at approximately $B/A\approx 0.5$.  These results establish a phase transition from the paired CF state into the CF Fermi sea as a function of the range of the three body interaction, which appears continuous to the extent we can surmise from our finite size study.  We note that we have shown here (and below) results only for the largest system that we have studied for each filling factor, but the smaller systems are fully consistent with our conclusions.

The formation of composite fermions at $\nu=1/2$ suggests the possibility of other phenomenology associated with composite fermions. We first study the states of $H_{\rm 3-body}$ at several fractions of the form $\nu=n^*/(2n^*\pm 1)$ and $\nu=1-n^*/(2n^*\pm 1)$. Fig.~\ref{overlap}(c) depicts the ground state overlap as a function of $B/A$, approaching a very high maximum at what we term the ``optimal" value of $B/A$.  Fig.~\ref{spectra} again illustrates a drastic reorganization of the low energy spectrum of $H_{\rm 3-body}$ as $B/A$ is turned on.  It is evident that near the optimal $B/A$, all states studied here (except $\nu=1/2$) are incompressible, in that they have a uniform ($L=0$) ground state separated from excitations by a robust gap.  Furthermore, the spectrum bears a striking resemblance to the weakly interacting  CF spectrum, shown in the bottom panel, not only for the $L=0$ ground state but also for the low energy branch of neutral excitations. The overlaps confirm the ground state as the CF-IQHE state and the low energy excitations as the CF excitons.  

\begin{figure}
\begin{center}
\includegraphics[width=\columnwidth]{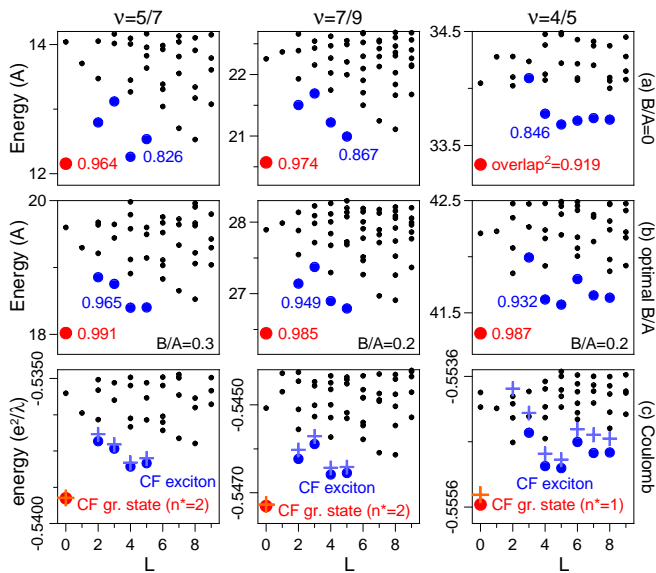}
\end{center}
\caption{\label{spectra2} (Color online) Emergence of composite fermions dressed with four vortices. Same as in Fig.~\ref{spectra}, but for filling factors $\nu=5/7$, 7/9 and 4/5 for $N=$19, 23 and 28, respectively. For $\nu=4/5$ the CF exciton state at $L=2$ is not identified in the exact spectrum, and also not used for the overlap calculation, because it merges into the continuum of two-exciton states at small $L$. 
}
\end{figure}

Figs.~\ref{overlap}(d)  and \ref{spectra2} demonstrate that the three body interaction not only creates composite fermions carrying two vortices ($^2$CFs) in the region $2/3\geq \nu >2/5$ but also composite fermions carrying four vortices ($^4$CFs) for $\nu>2/3$ at fillings of the form $\nu=1-n^*/(4n^*\pm 1)$. In fact, $^4$CFs are more robust;  for $\nu<2/3$ it requires a nonzero $B/A$ to create weakly interacting composite fermions, whereas for $\nu\geq 2/3$ the CF physics is established already at $B/A=0$. 

We have also studied numerous systems in between the special filling factors shown above, and in all cases we find that for a range of $B/A$ values the physics is consistent with weakly interacting composite fermions.  Specifically, the low energy band of states is well described, qualitatively and quantitatively, in terms of CF quasiparticles or CF quasiholes on top of a CF-IQHE state.

The three body interaction does not respect particle hole symmetry in the lowest LL, which is responsible for the qualitatively distinct physics for $\nu<2/5$ and $3/5<\nu<1$, and also for the Pfaffian state, which is not particle-hole symmetric. A surprising outcome of this work is that for a range of $B/A$ values, the three body interaction behaves similarly as the two body Coulomb interaction insofar as the low energy physics of the correlated states is concerned, implying a partial restoration of the particle hole symmetry in the region $3/5>\nu>2/5$ when the longer range part of the three body interaction is turned on.

\begin{figure}[t!]
\begin{center}
\includegraphics[width=\columnwidth]{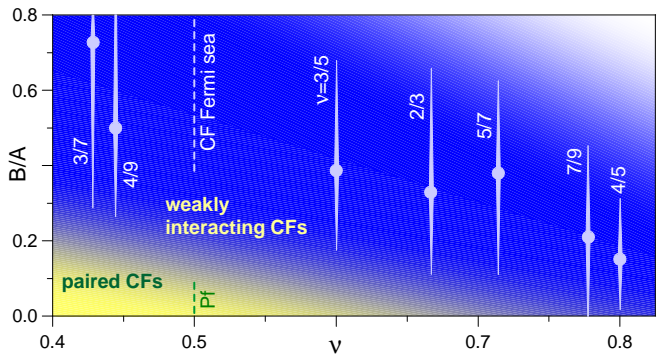}
\end{center}
\caption{\label{phase-diagram} (Color online) The phase diagram of FQHE for three body interaction.  The region where the physics is described in terms of weakly interacting composite fermions is shaded in blue, and the region of paired composite fermions in yellow.  At $\nu=1/2$ the Pfaffian FQHE state undergoes a transition into the CF Fermi sea with increasing $B/A$. Each vertical line indicates the range where squared overlap of the exact ground state of the three body interaction with the weakly interacting  CF state exceeds 90\% (for our largest system), with the dot marking the position of the maximum overlap. The phase boundaries are to be viewed as being semi-quantitative.
}
\end{figure}

Our principal conclusion is summarized in the phase diagram in Fig.~\ref{phase-diagram}. The emergence of composite fermions and the FQHE for three body interactions is unexpected.  The canonical model for FQHE is that of pairwise interaction with a strong short range repulsion, and composite fermions are thought to materialize to minimize the short range part of the two body interaction. This is most evident from the fact that the ``unprojected" wave functions of composite fermions \cite{Jain89} vanish much faster than what is required by the Pauli principle when {\em two} particles come close.  Nevertheless, the appearance of almost free composite fermions enables an understanding of the physics of lowest LL fermions with three body interaction at a level that is almost as detailed and accurate as that available for the Coulomb interaction. 

Finally, although our study deals with fermions, experience from earlier work \cite{Cooper99,Chang05} suggests that for appropriate three body interactions bosons will also composite-fermionize by capturing a single quantized vortex (not to be confused with the vortex in the order parameter field of the Bose-Einstein condensate), to produce FQHE at $\nu=n^*/(n^*\pm 1)$ and a CF Fermi sea or a paired state at $\nu=1$.  We have not yet investigated this possibility.

We are grateful to Nigel Cooper and Steven Simon for valuable discussions. A.W. acknowledges the support of the EU under the Marie Curie grant PIEF-GA-2008-221701. C.T. was supported by Science, Please! Innovative Research Teams, SROP-4.2.2/08/1/2008-0011.  We thank Information Systems Services at Lancaster University for their assistance with the High Performance Cluster, where part of the computation was performed.

\end{document}